\documentclass[fleqn,usenatbib]{mnras}

\usepackage{newtxtext,newtxmath}

\usepackage[T1]{fontenc}

\DeclareRobustCommand{\VAN}[3]{#2}
\let\VANthebibliography\thebibliography
\def\thebibliography{\DeclareRobustCommand{\VAN}[3]{##3}\VANthebibliography}

\usepackage{graphicx}	
\usepackage{amsmath}	
\usepackage[caption=false]{subfig} 
\usepackage{pdflscape}	
\usepackage[referable]{threeparttablex} 
\usepackage{booktabs} 
\usepackage[normalem]{ulem}

\newcommand{\nodata}{ ~$\cdots$~ } 

\title[Dust Grains in Young L Dwarf Atmospheres Are Heavier]{Ultracool Dwarfs Observed with the Spitzer Infrared Spectrograph - III. Dust Grains in Young L Dwarf Atmospheres Are Heavier}

\author[Su\'arez \& Metchev]{
Genaro Su\'arez,$^{1,2}$\thanks{E-mail: gsuarez@amnh.org}
Stanimir Metchev,$^{2,3}$
\\
$^{1}$Department of Astrophysics, American Museum of Natural History, Central Park West at 79th St, New York, NY, 10024, USA\\
$^{2}$Department of Physics and Astronomy, The University of Western Ontario, 1151 Richmond St, London, Ontario, N6A 3K7, Canada\\
$^{3}$Institute for Earth and Space Exploration, The University of Western Ontario, 1151 Richmond St, London, Ontario, N6A 3K7, Canada\\
}

\date{Accepted XXX. Received YYY; in original form ZZZ}

\pubyear{2015}

\begin{document}
\label{firstpage}
\pagerange{\pageref{firstpage}--\pageref{lastpage}}
\maketitle

\definecolor{orange}{rgb}{0.98, 0.6, 0.01}

\begin{abstract}
Analysis of all archival 5--14~micron spectra of field ultracool dwarfs from the Infrared Spectrograph on the Spitzer Space Telescope has shown that absorption by silicates in the 8--11~micron region is seen in most L-type (1300~K to 2200~K) dwarfs. The absorption is caused by silicate-rich clouds in the atmospheres of L dwarfs and is strongest at L4--L6 spectral types. Herein we compare averages of the mid-infrared silicate absorption signatures of L3--L7 dwarfs that have low ($\lesssim$10$^{4.5}$~cm~s$^{-2}$) vs.\ high ($\gtrsim$10$^5$~cm~s$^{-2}$) surface gravity. We find that the silicate absorption feature is sensitive to surface gravity and indicates a difference in  grain size and composition between dust condensates in young and old mid-L dwarfs. The mean silicate absorption profile of low-gravity mid-L dwarfs matches expectations for $\sim$1~micron-sized amorphous iron- and magnesium-bearing pyroxene (Mg$_x$Fe$_{1-x}$SiO$_3$) grains. High-gravity mid-L dwarfs have silicate absorption better represented by smaller ($\lesssim$0.1~$\mu$m) and more volatile amorphous enstatite (MgSiO$_3$) or SiO grains. This is the first direct spectroscopic evidence for gravity-dependent sedimentation of dust condensates in ultracool atmospheres. It confirms theoretical expectations for lower sedimentation efficiencies in low-gravity atmospheres and independently confirms their increased dustiness.
\end{abstract}




\section{Introduction}
\label{sec:introduction}
The atmospheres of L spectral type dwarfs have effective temperatures between approximately 1300--2200~K \citep{Kirkpatrick2005,Filippazzo_etal2015}. It is observationally established, based both on their very red optical to near-infrared colours \citep{Knapp_etal2004,Kirkpatrick2005} and on the direct detection of silicates in mid-infrared spectra \citep{Roellig_etal2004,Cushing_etal2006,Looper_etal2008b,Suarez-Metchev2022}, that these atmospheres are dusty. Theoretically, silicon-bearing molecules such as quartz (SiO$_2$), pyroxene (Mg$_x$Fe$_{1-x}$SiO$_3$), and olivine (Mg$_x$Fe$_{2-x}$SiO$_4$), including the Mg-rich endmembers enstatite (MgSiO$_3$) and forsterite (Mg$_2$SiO$_4$), are expected to start forming 0.1--1~$\mu$m-sized silicate dust condensates at $\leq$2000 K \citep{Burrows_etal1997,Burrows_etal2001,Ackerman-Marley2001,Lodders2002}. The theory dictates that the silicate condensate clouds sediment to deeper levels as the L-type substellar atmosphere cools, and at $\lesssim$1300~K effective temperatures the clouds drop below the altitude of the $\sim$0.1~bar pressure level of the mid-infrared photosphere \citep{Luna-Morley2021}. The onset of dust cloud formation at warm temperatures and their later sedimentation at cooler temperatures have been observationally confirmed from 5--14~$\mu$m mid-infrared spectra of L dwarfs \citep{Suarez-Metchev2022}.

In addition to being dependent on temperature, dust sedimentation in the atmosphere is also a function of surface gravity. Cloud settling is less efficient at lower gravity, which should lead to increased cloud vertical depth and optically thicker cloud decks \citep{Stephens_etal2009,Madhusudhan_etal2011,Marley_etal2012,Charnay_etal2018}. Indeed, unusually red optical/near-infrared colours are a well-known trait of young L dwarfs (e.g., \citealt{Kirkpatrick_etal2006,Kirkpatrick_etal2008,Looper_etal2008b,Faherty_etal2016}), which have lower surface gravity because of their larger radii at their earlier evolutionary stages (e.g., \citealt{Burrows_etal1997}). Indirect evidence of the greater importance of dust in younger L-type atmospheres also comes from their enhanced brightness variations \citep{Metchev_etal2015,Biller_etal2015,Vos_etal2018,Vos_etal2019,Bowler_etal2020}, understood to be caused by the rotationally modulated visibility of large dust cloud features \citep{Burgasser_etal2002b,Artigau_etal2009}. 

The role of gravity in sedimentation also dictates that heavier dust grains, made of potentially more refractory compounds, should sediment first, leaving lighter grains at higher altitudes \citep{Ackerman-Marley2001,Cooper_etal2003}.  However, this has yet to be observationally demonstrated, and is the main finding of the present study.

The physical properties and mineral composition of silicate dust in brown dwarf atmospheres remain relatively unexplored. They require a population analysis of high signal-to-noise ratio (SNR) mid-infrared spectra of these objects that awaits observations with the James Webb Space Telescope (JWST). To date, only a limited sample has been analysed based on low-resolution spectra with the Spitzer Space Telescope Infrared Spectrograph (IRS; \citealt{Houck_etal2004}), and only during Spitzer's 2003--2009 Cryogenic Mission. Namely, theoretical modelling to match six moderate- to high-SNR L dwarf Spitzer IRS spectra \citep{Roellig_etal2004,Cushing_etal2006,Looper_etal2008b} has shown that dust clouds in L dwarfs comprise a combination of amorphous silicates (quartz, enstatite, and forsterite) and iron condensed in $\lesssim$1~$\mu$m-sized grains \citep{Cushing_etal2006,Burningham_etal2021,Luna-Morley2021}. For over a decade, this small sample of L dwarf spectra has been the only source to probe cloud chemistry directly. Recently, silicate absorption was also detected in the atmosphere of the late-L planetary-mass companion VHS~J125601.92$-$125723.9~b using JWST Mid-Infrared Instrument (MIRI) spectra \citep{Miles_etal2022} and forsterite clouds have been inferred in the IRS spectra of two young ($\sim200$ Myr) early-T dwarf atmospheres \citep{Vos_etal2023}.

The study of \citet{Suarez-Metchev2022} re-assessed all 113 Spitzer IRS observations of $\geq$M5 ($T_{\rm eff} \leq 3000$~K) dwarfs outside of known $\lesssim$10~Myr-old star-forming regions. A comparative analysis of the 69 L dwarf spectra in the sample established many more instances of spectroscopic silicate cloud detections beyond the six known prior. The study revealed that silicate clouds first form, then thicken, and sediment out of the visible atmosphere as the effective temperature decreases from $\approx$2000 K to $\approx$1300 K. While the \citet{Suarez-Metchev2022} analysis focused mostly on $\gtrsim$500~Myr-old field dwarfs outside of known star-forming regions, it also included four L3--L6.5-type dwarfs with spectroscopic and kinematic evidence of youth (few tens of Myr). This sample of relatively young objects, complemented with old counterparts, allows an investigation of the surface gravity (or age) dependence of silicate absorption in ultracool atmospheres.
 
Herein we compare the silicate absorption in the mid-infrared Spitzer IRS spectra of young, low-surface gravity mid-L dwarfs ($\log g \lesssim 4.5$; 1800~K $\gtrsim T_{\rm eff} \gtrsim$ 1200~K) to their older, higher-gravity ($\log g{\gtrsim}5.0$) counterparts in the Spitzer IRS sample. Our analysis confirms the theoretical expectations of gravity-dependent sedimentation of dust clouds from \citet{Ackerman-Marley2001,Cooper_etal2003}. Specifically, we find that heavier dust grains have higher abundances in lower-gravity L-type atmospheres than they do at higher surface gravity. In Section~\ref{sec:methods} we describe the sample and the approach followed to compare silicate absorption profiles in young and old ultracool atmospheres. An investigation of how the silicate absorption depends on surface gravity is presented in Section~\ref{sec:results}. In Section~\ref{sec:discussion} we compare observed silicate absorption profiles to theoretical expectations. We summarise our conclusions in Section~\ref{sec:summary}.

\section{Sample and Methods}
\label{sec:methods}

To assess the surface gravity dependence of the silicate absorption feature we define sub-samples of young and old mid-L dwarfs with the strongest silicate features. We then use a separate sub-sample of mid-L dwarfs with no discernible silicate features to estimate the continuum for the silicate-rich spectra. Table~\ref{tab:dwarfs} lists the L3--L7 dwarfs in each sub-sample and their relevant properties.

\subsection{Selection of Mid-L Dwarfs with Silicate-rich and Silicate-poor Spectra}
\label{sec:sample_selection}
\begin{figure}
	\centering
	\includegraphics[width=1.0\linewidth]{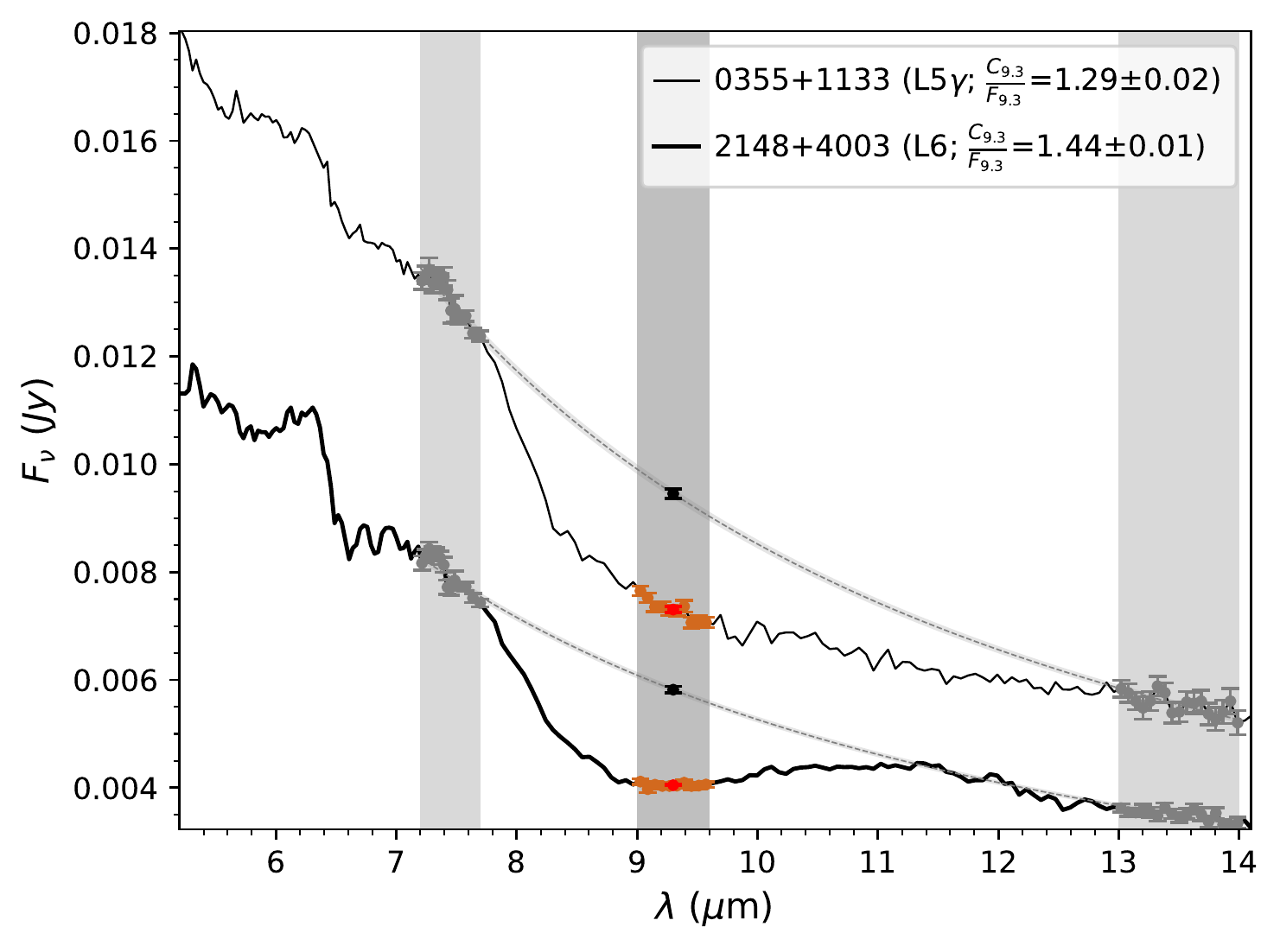}
	\caption{Silicate index measurements for the low-surface gravity ($\log g=4.45\pm0.21$; \citealt{Filippazzo_etal2015}) L5$\gamma$ dwarf 2MASS J03552337+1133437 (top spectrum) and the high-surface gravity ($\log g=5.13\pm0.28$; \citealt{Filippazzo_etal2015}) L6 dwarf 2MASS J21481628+4003593 (bottom spectrum). The silicate index is defined as the ratio of the interpolated continuum (black point; $C_{9.3}$) to the median flux (red point; $F_{9.3}$) at 9.3~$\mu$m in a 0.6~$\mu$m-wide window (brown points in the central dark grey vertical region). The interpolated continuum (black dashed curve with uncertainties shown by the grey shaded region) is obtained as a linear $\log F_\nu$ vs.\ $\log\lambda$ fit to the fluxes in the 7.2--7.7 $\mu$m and 13--14~$\mu$m continuum regions (grey points in the outer light grey vertical regions). The $C_{9.3}/F_{9.3}$ silicate index is indicated in the legend.}
	\label{fig:silicate_index_example}
\end{figure}

To measure silicate absorption strengths we iterated on \citet{Suarez-Metchev2022}, and defined a silicate index as the ratio of an interpolated continuum to the median flux at 9.3~$\mu$m in a 0.6~$\mu$m-wide window. The interpolated continuum was calculated as the best $\log F_\nu$ vs.\ $\log \lambda$ linear fit to the fluxes in two regions that frame the silicate feature and are mostly free of other molecular absorption: 7.2--7.7~$\mu$m and 13.0--14.0~$\mu$m (Figure~\ref{fig:silicate_index_example}). We changed the central wavelength of the silicate absorption index (to 9.3~$\mu$m from 9.0~$\mu$m) compared to \citet{Suarez-Metchev2022} to better match theoretical expectations from \citet{Luna-Morley2021}. The long-wavelength continuum region was further moved to 13.0--14.0~$\mu$m vs.\ 11.2--11.8~$\mu$m to exclude as much as possible of the red wing of the silicate feature for both young and old dwarfs (Figure~\ref{fig:silicate_index_example}).

\begin{figure*}
	\centering
	\subfloat[]{\includegraphics[width=.5\linewidth]{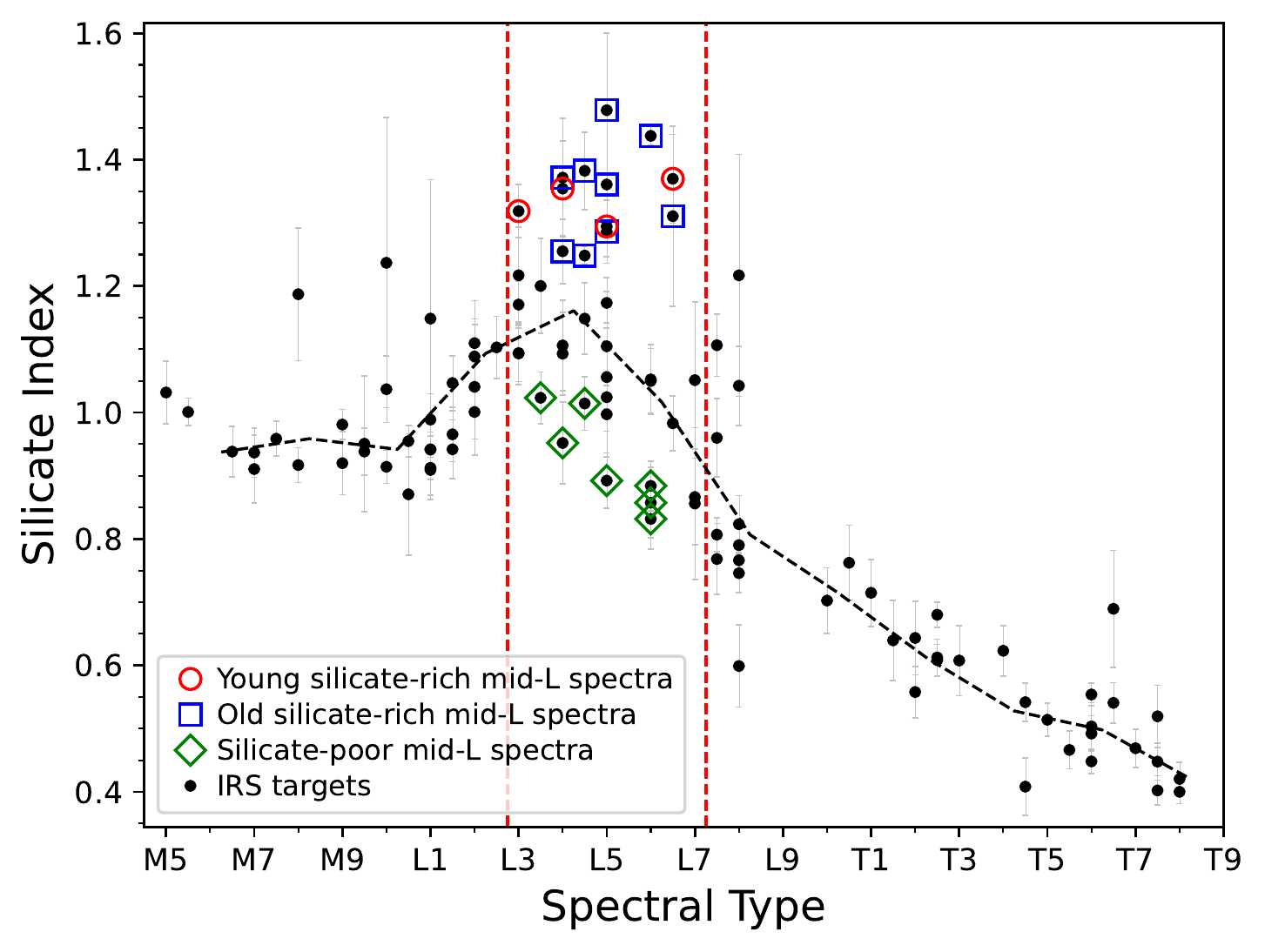}
	\label{fig:silicate_index_vs_spt_a}}
	\subfloat[]{\includegraphics[width=.5\linewidth]{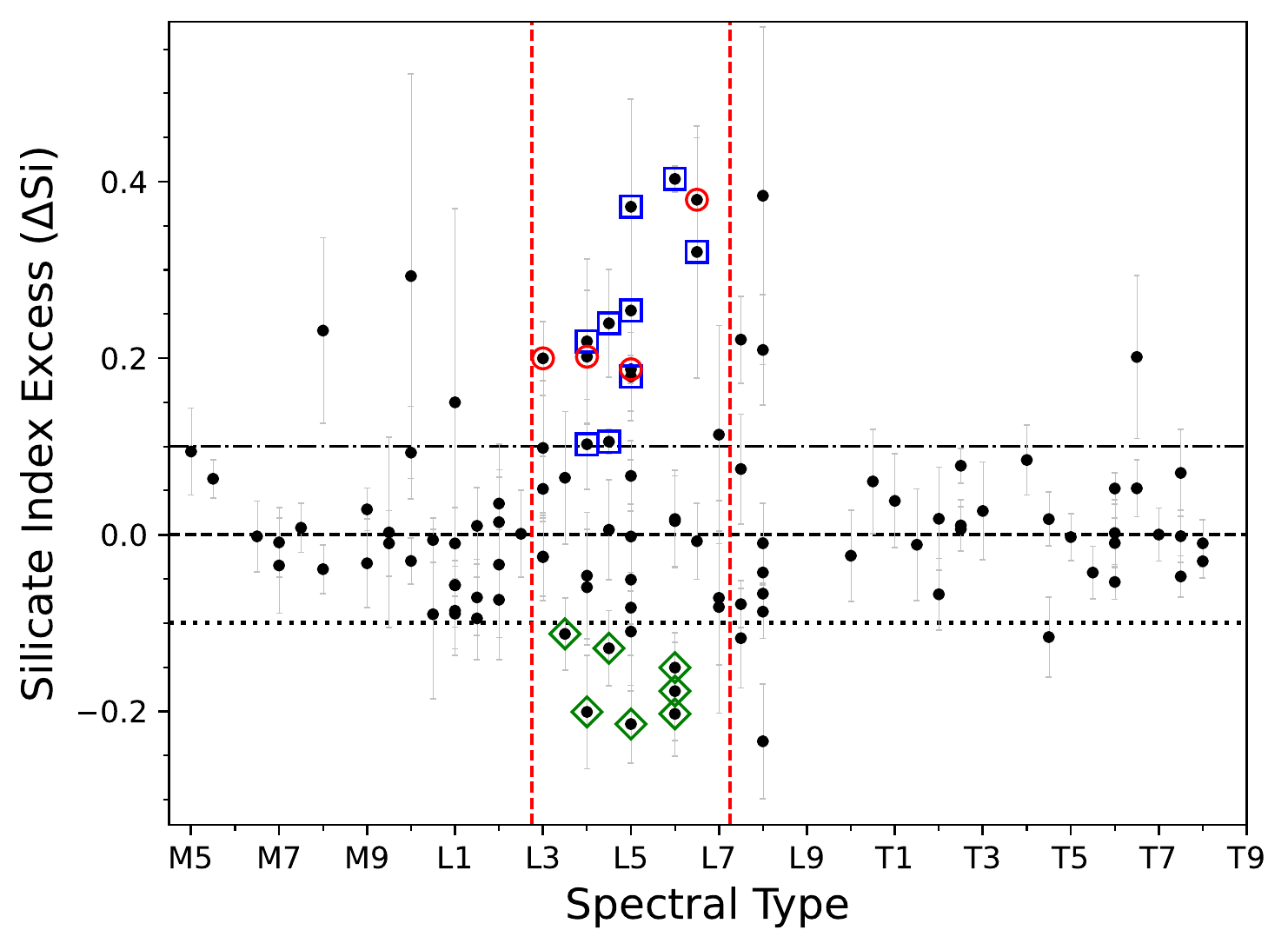}
	\label{fig:silicate_index_vs_spt_b}}
	\caption{\textbf{(a):} Silicate index as a function of spectral type for all 113 M5--T9 dwarfs with IRS spectra (black points) in \citet{Suarez-Metchev2022}. The black dashed curve corresponds to the median silicate index in bins of two spectral subtypes. The vertical red dashed lines delimit the L3--L7 spectral type range (slightly shifted for clarity) within which we analyse young objects with silicate-rich spectra (red circles), old objects with silicate-rich spectra (blue squares), and objects with silicate-poor spectra (green diamonds). \textbf{(b):} Silicate index excess ($\Delta$Si), defined as the difference between the measured and the median (black dashed line) silicate index for each object (Section~\ref{sec:sample_selection}). $\Delta{\rm Si}>0.10$ (above the black dash-dotted line) and $\Delta{\rm Si}<-0.10$ (below the black dotted line) were used to select silicate-rich and silicate-poor spectra. The rest of the symbols are the same as in (a).}
	\label{fig:silicate_index_vs_spt}
\end{figure*}

Figure~\ref{fig:silicate_index_vs_spt_a} shows the updated silicate index measurements as a function of spectral type for the M5--T8 dwarfs in \citet{Suarez-Metchev2022}. We adopt optical spectral types for M and L dwarfs and infrared types for T dwarfs. The dashed line traces the median silicate index value in two spectral subtype-wide bins. As observed in \citet{Suarez-Metchev2022}, the L dwarfs show a systematic excess in silicate index values above a declining trend from late-M to T dwarfs. This indicates near-ubiquitous 8--11~$\mu$m silicate absorption in L dwarfs. The updated definition of the silicate index with a redder central wavelength and a wider continuum span compared to \citet{Suarez-Metchev2022} shows that silicate absorption is already detectable in the average L1 dwarf, slightly earlier than the L2 onset noted in \citet{Suarez-Metchev2022}.

We define a silicate index excess ($\Delta$Si) for each ultracool dwarf as the difference between its measured silicate index and the median index for its spectral type (black dashed curve in Figure~\ref{fig:silicate_index_vs_spt_a}). Figure~\ref{fig:silicate_index_vs_spt_b} shows the silicate excess as a function of spectral type for the full sample in \citet{Suarez-Metchev2022}. We constrain our analysis to L3--L7 dwarfs, as they have the strongest absorption on average, and their 8--11~$\mu$m silicate absorption is not contaminated by the 7--9~$\mu$m methane absorption that sets in at spectral type L8 \citep{Suarez-Metchev2022}. From all L3--L7 dwarfs, we select those with $\Delta {\rm Si}>0.10$ and for which $\Delta {\rm Si}$ is positive by more than twice its uncertainty (i.e., SNR ($\Delta {\rm Si})>2$) as mid-L dwarfs with silicate-rich spectra. This selection includes three of the five mid-L dwarfs previously known or suspected to have silicate absorption: 2MASS J18212815+1414010 (L4.5), 2MASS J21481633+4003594 (L6), and 2MASS J22244381$-$0158521 \citep[L4.5;][]{Cushing_etal2006,Looper_etal2008b}. The remaining two---2MASS J00361617+1821104 (L3.5) and 2MASS J15074769$-$1627386 \citep[L5;][]{Cushing_etal2006}---have silicate indices closer to the median silicate index for their spectral subtypes. That is, they still likely have silicate absorption, but it does not stand out in strength compared to other L dwarfs of similar spectral types. (The sixth L dwarf with previously reported silicate absorption is 2MASS J02550357$-$4700509 \citep[L8;][]{Roellig_etal2004}, which is outside of the L3--L7 spectral type range considered here.)

We select seven objects in the L3--L7 range with the most negative silicate excess as a sub-sample representative of mid-L dwarfs with silicate-poor spectra. Mirroring the selection of silicate-rich spectra, we select these to have $\Delta {\rm Si} < -0.10$ and a $\Delta {\rm Si}$ value that is negative by more than twice its uncertainty. It is highly likely that the atmospheres of the mid-L dwarfs with silicate-poor spectra are still abundant in silicates. However, the silicate condensates may have mostly settled to a layer below the visible 8--11~$\mu$m photosphere. Hence, their spectra are representative of the expected mid-infrared continuum and absorption by other non-silicate species, such as water and methane.

\subsection{Assignment of Surface Gravity}
\label{sec:logg}

We use spectral features known to be associated with gravity, corroborated with kinematic association, or lack thereof, to known young stellar moving groups to assign mid-L dwarfs into categories of low- versus high-surface gravity. We disfavour surface gravity determinations from photospheric model fits to the low-resolution Spitzer IRS spectra themselves, or to broad spectral energy distributions (SEDs) that are not constrained by age. Gravity determinations from Spitzer IRS spectra do not conform well with results from near-infrared spectra that contain pressure-sensitive lines or features \citep[e.g.,][]{Suarez_etal2021a}. Broad SEDs are excellent for bolometric luminosity estimates, but are poor for determining surface gravities \citep{Cushing_etal2008}. They benefit from an evolutionary model-dependent radius or mass estimate for improved effective temperature and surface gravity determination \citep[e.g.,][]{Cushing_etal2008, Filippazzo_etal2015}.

There are four L3--L6.5 dwarfs among the 13 silicate-rich spectra with a $\gamma$ spectroscopic designation indicative of low surface gravity and hence youth. We extract these as a sub-sample of young silicate-rich mid-L spectra. The remaining nine (L4--L6.5) dwarfs with silicate-rich spectra have no previously published spectroscopic indications of youth. We assign these to the sub-sample of old silicate-rich mid-L spectra. We verify the age classification of our mid-L dwarfs with silicate-rich spectra by checking whether any of them may be kinematically associated with known nearby young stellar moving groups. Three of the $\gamma$-classified objects (2MASS J03552337+1133437, 2MASS J14252798$-$3650229, and 2MASS J22443167+2043433) are bona fide members of the 110--150~Myr-old \citep{Luhman_etal2005b,Barenfeld_etal2013} AB Doradus moving group \citep{Liu_etal2013,Gagne_etal2015,Vos_etal2018} and 2MASS J05012406$-$0010452 is a possible member of the Columba or Carina moving groups \citep{Gagne_etal2015}, which are coeval at 20--40~Myr \citep{Torres_etal2008}. Conversely, the BANYAN $\Sigma$ algorithm \citep{Gagne_etal2018b} tool indicates that seven of the nine old silicate-rich mid-L dwarfs have a $>$96\% probability of being field objects. Only 2MASS J17534518$-$6559559 (L4) and 2MASS J09201223+3517429 (L6.5) have 85\% and 14\% probabilities of membership in the 40--50~Myr-old \citep{Zuckerman2019} Argus and 110--150~Myr AB Doradus moving groups, respectively. However, kinematic association with a young stellar moving group is not a sufficient indicator of youth. Because of the lack of spectroscopic confirmation of low surface gravity, we keep these two objects in our sample of old mid-Ls. Their IRS spectra have low-SNR ($\approx$10 at 9~$\mu$m), and so do not significantly contribute to our representation of old silicate-rich mid-L dwarfs (Section~\ref{sec:silicate_scaling}).

The above division in low vs.\ high surface gravity sub-samples is mostly consistent with surface gravity determinations for the individual objects in the published literature. The four young dwarfs all have $\log g\lesssim4.5$ ($g\lesssim10^{4.5}$~cm~s$^{-2}$) from combined constraints on their bolometric luminosities from trigonometric parallaxes and (photospheric model fits to) SEDs, and on the ages for the respective parent associations \citep{Filippazzo_etal2015,Gagne_etal2015,Vos_etal2018}. The old dwarfs with gravity estimates all have $\log g\gtrsim5.0$ ($g\gtrsim10^{5.0}$~cm~s$^{-2}$) \citep{Cushing_etal2008,Gagne_etal2015,Filippazzo_etal2015,Dupuy-Liu2017,Luna-Morley2021,Burningham_etal2021}. Some studies point to higher $\log g$ values for two of the four young dwarfs (2MASS J22443167+2043433 in \citealt{Stephens_etal2009}; 2MASS J14252798$-$3650229 in \citealt{Gagne_etal2015}), or to a lower $\log g$ for one of the old dwarfs \citep[2MASS J22244381$-$0158521;][]{Cushing_etal2008,Stephens_etal2009}. All of the discrepant gravities were obtained by fitting atmospheric models to broad SEDs without an age constraint, and so are potentially less accurate. For instance, while \citet{Gagne_etal2015} find a field-like $\log g=5.0\pm0.5$ for 2MASS J14252798$-$3650229 from an age-unconstrained photospheric model to the SED, they also note that this object is a young $\gamma$-class L4 dwarf from analysis of narrower spectral regions sensitive to surface gravity. Similarly, \citet{Cushing_etal2008} and \citet{Stephens_etal2009} find a relatively low $\log g=4.5\pm0.5$ from age-unconstrained photospheric model fits to the SED of 2MASS J22244381$-$0158521. However, when combined with an evolutionary model appropriate for its otherwise field-age kinematics, the photospheric fit to the SED results in a higher surface gravity \citep[$\log g\ge5.2$;][]{Cushing_etal2008, Filippazzo_etal2015}.

Figure~\ref{fig:silicate_index_vs_spt} highlights the selected sub-samples of mid-L dwarfs from Table~\ref{tab:dwarfs} according to their silicate strength and surface gravity.

\begin{figure*}
	\centering
	\subfloat[]{\includegraphics[width=.5\linewidth]{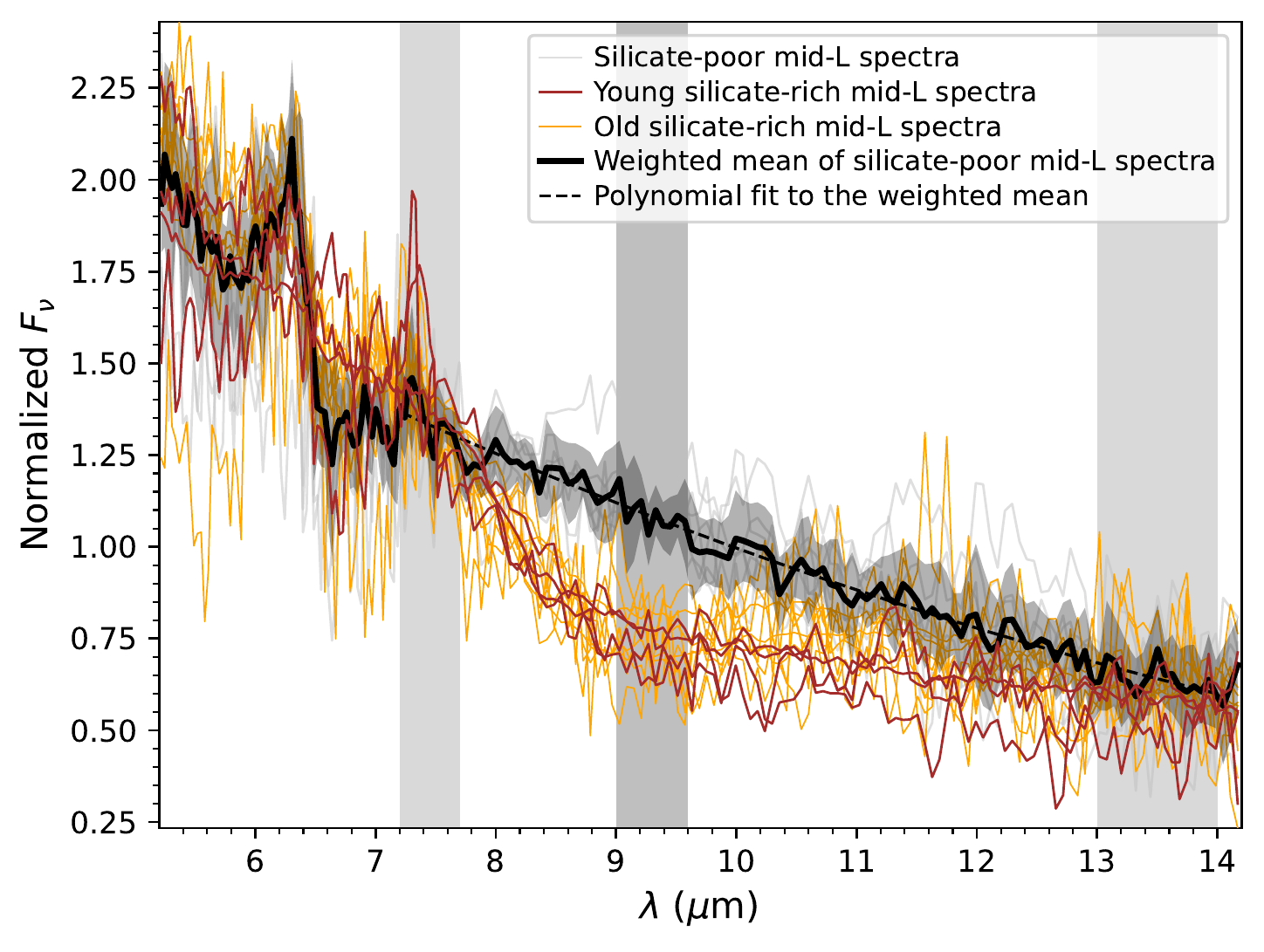}
	\label{fig:silicate_absorption_a}}
	\subfloat[]{\includegraphics[width=.5\linewidth]{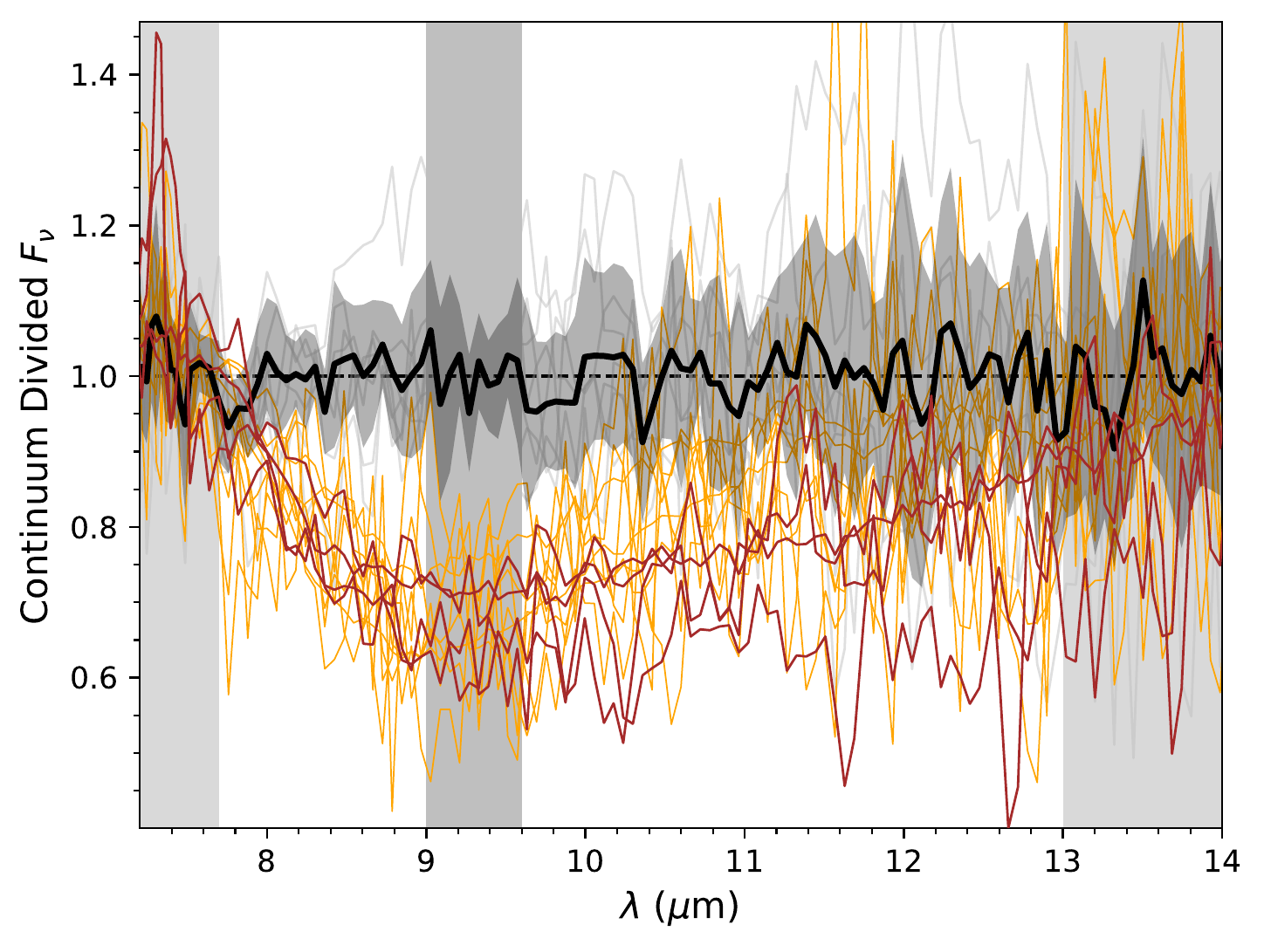}
	\label{fig:silicate_absorption_b}}
	\caption{\textbf{(a):} Normalised comparison of four young silicate-rich (brown curves), seven old silicate-rich (orange curves), and seven silicate-poor (grey curves) L3--L6.5 spectra. The normalisation was performed to the median flux in the 7.2--7.7~$\mu$m and 13.0--14.0~$\mu$m continuum windows (light grey vertical regions) that surround the silicate absorption. The 9.0--9.6~$\mu$m window (medium grey vertical region) denotes the range where silicate absorption is on average the strongest. The weighted average of the silicate-poor spectra is indicated by the black curve and its uncertainty by the surrounding dark grey shaded region. The best second-order polynomial fit to this combined spectrum in the 7.2--14.0~$\mu$m range is represented by the black dashed curve. \textbf{(b):} The 7.2--14.0~$\mu$m sub-range of the same spectra as in (a), divided by the polynomial fit to the average silicate-poor mid-L continuum. By construction, the continuum-divided weighted average (black curve) of silicate-poor spectra scatters around unity (dashed line; same as in (a)).}
	\label{fig:silicate_absorption}
\end{figure*}

\subsection{Continuum Removal}
\label{sec:continuum_removal}

Figure~\ref{fig:silicate_absorption_a}  illustrates the difference in the silicate absorption feature between the selected sub-samples of silicate-poor and silicate-rich mid-L spectra. The full 5.2--14.2~$\mu$m spectra are shown scaled to the same average flux in the 7.2--7.7~$\mu$m and 13.0--14.0~$\mu$m continuum regions. The weighted average of the silicate-poor mid-L spectra (black curve), obtained using the inverse squared values of the errors on the scaled flux as weights, traces the mid-infrared continuum in the near-total absence of silicate absorption. It does not exhibit a significant scatter and is well-represented by a second-order polynomial fit (black dashed curve) between 7.2--14.0~$\mu$m. The silicate-rich spectra exhibit characteristic broad absorption from 8~$\mu$m out to 11--13~$\mu$m.

To isolate the silicate absorption signature in silicate-rich mid-L dwarfs from that of other potential continuum absorbers, we divide all normalised IRS spectra by the quadratic fit to the weighted average silicate-poor mid-L continuum. The continuum-divided mid-L spectra are shown in Figure~\ref{fig:silicate_absorption_b}, where we have narrowed the rendering to the 7.2--14.0~$\mu$m range that spans the $\sim$8--13~$\mu$m silicate absorption. The underlying assumption for the continuum removal is that any non-silicate absorbers (such as water or methane) are present at similar strength in the silicate-poor and silicate-rich mid-L spectra. While that appears to mostly be the case, given the similarity among the continuum-divided mid-L spectra outside the silicate absorption region (Figure~\ref{fig:silicate_absorption_b}), there remain some systematic differences. Namely, the spectra of mid-L dwarfs with enhanced silicate absorption generally have higher fluxes at $<$7.5~$\mu$m and lower fluxes at $\gtrsim$13~$\mu$m compared to the silicate-poor spectra. The culprit for the short-wavelength difference is likely water absorption. Mid-L dwarfs with higher silicate content in the upper atmosphere likely have some of the gas-phase water column blocked, and hence exhibit weaker water absorption. At long wavelengths the culprit is likely silicates themselves. The absorption signatures of the nine-micron amorphous pyroxene (including enstatite) features may extend beyond 13~$\mu$m \citep{Luna-Morley2021}. 

Despite the systematics at the edges, the continuum-divided silicate-rich spectra do isolate the main part of the silicate feature well. A difference in the red wing of this feature is discernible in Figure~\ref{fig:silicate_absorption_b} between the young (brown) and old (orange) silicate-rich mid-L spectra, with the young mid-L spectra consistently tracking below the old spectra at $>$10~$\mu$m. However, most of the individual spectra are too noisy to assess this  difference. We proceed by scaling and averaging the silicate absorption profiles themselves. 

\subsection{Scaling and Averaging of Silicate Absorption Profiles}
\label{sec:silicate_scaling}
\begin{figure*}
	\centering
	\subfloat[]{\includegraphics[width=.5\linewidth]{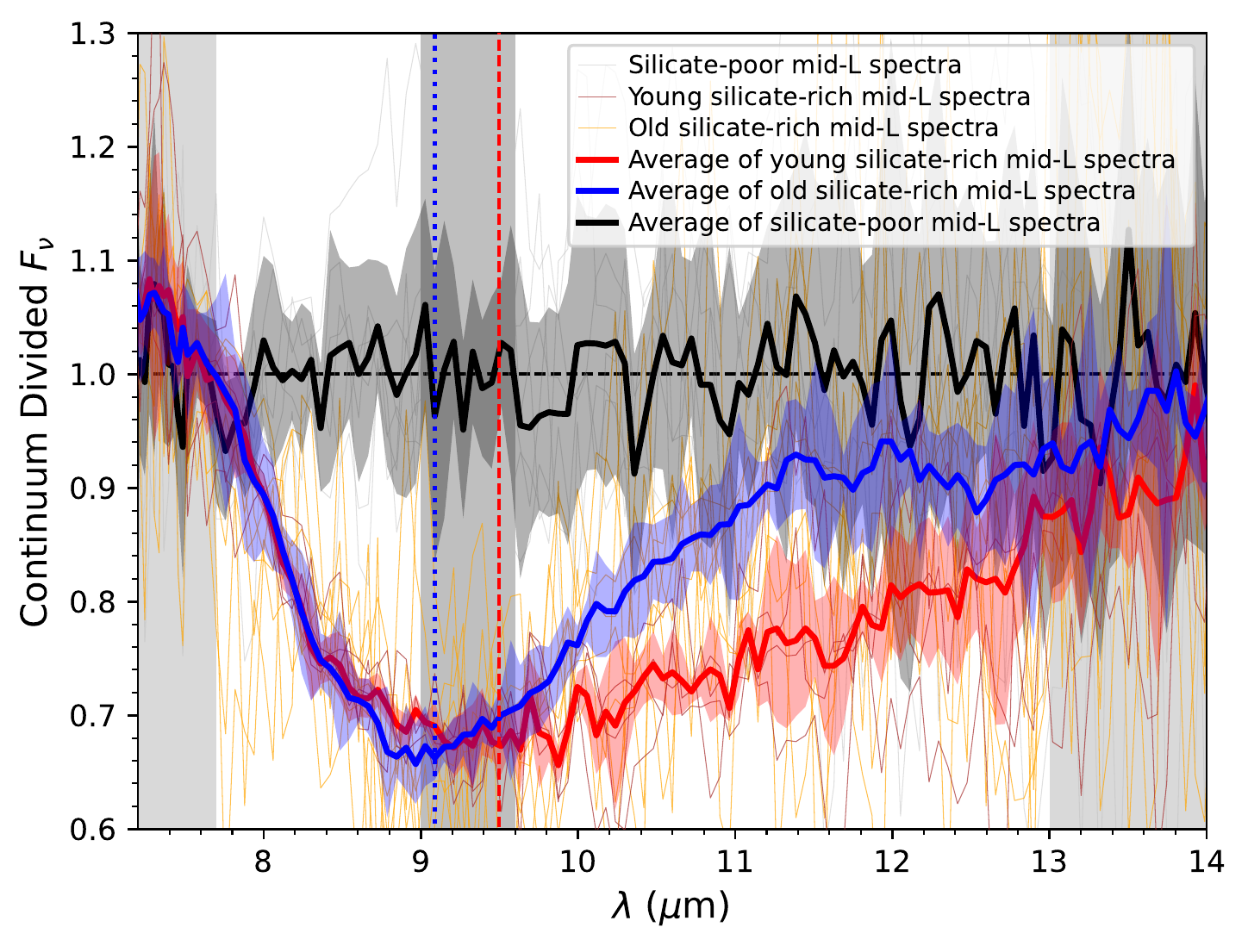}
	\label{fig:silicate_absorption_comparison_a}}
	\subfloat[]{\includegraphics[width=0.5\textwidth]{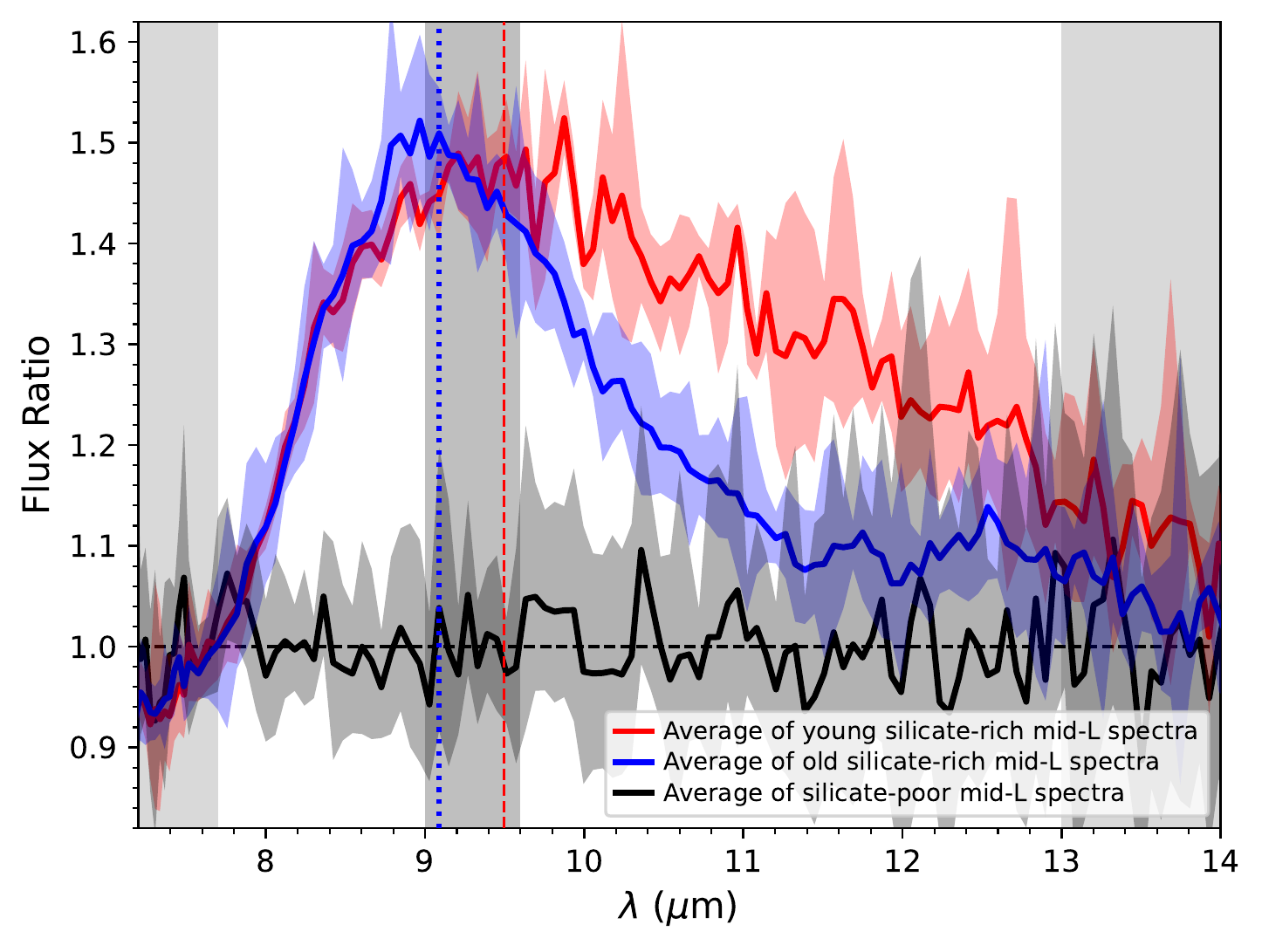}
	\label{fig:silicate_absorption_comparison_b}}
	\caption{\textbf{(a)} Continuum-divided 7.2--14.0~$\mu$m subsections of the mid-L dwarf IRS spectra from Figure~\ref{fig:silicate_absorption_b} with the strength of the silicate absorption in  silicate-rich spectra scaled to the same median flux in a 0.6~$\mu$m-wide window centred on 9.3~$\mu$m (vertical grey band). Weighted averages of these mid-L spectra are shown as follows: young silicate-rich spectra (red), old silicate-rich spectra (blue), silicate-poor spectra (black). The uncertainties of the weighted-average spectra are shown by shaded regions of the corresponding colour. The vertical red dashed and blue dotted lines mark the wavelengths of strongest silicate absorption in young and old silicate-rich mid-L spectra, respectively (see Section~\ref{sec:silicate_scaling}). Spectra of individual objects follow the same colour scheme as in Figure~\ref{fig:silicate_absorption}. \textbf{(b)} Inverse of fluxes in panel (a). This representation allows a direct comparison with flux ratio predictions for condensate cloud vs.\ grey cloud models in Figure 7 of \citet{Luna-Morley2021}.}
	\label{fig:silicate_absorption_comparison}
\end{figure*}

With the continua of the silicate-rich spectra factored out, we scale the strengths of the silicate absorption profiles to each other to account for differences in the silicate column density among objects. We use the flux median of the silicate-rich spectra within a 0.6~$\mu$m-wide region centred on 9.3~$\mu$m \citep{Cushing_etal2006,Luna-Morley2021} to scale all absorption profiles to the same depth (Figure~\ref{fig:silicate_absorption_comparison_a}). We can apply such scaling without fear of contamination from non-silicate absorbers, because we have already removed their effect after dividing by the silicate-poor continuum. In so far as the silicate species are similar among the different mid-L dwarfs, their normalised absorption profiles will also be similar as a function of wavelength, save for a multiplicative constant. However, small departures from similarities, as expected among different amorphous sub-types of silicates, are enhanced.

We take the weighted average of the scaled spectra for each of the sub-samples and obtain population-averaged silicate absorption profiles for young and old mid-L spectra. As for the combination of the silicate-poor spectra (Section~\ref{sec:continuum_removal}), we use the inverse squared values of the errors on the scaled flux as weights. The combined spectra are shown in Figure~\ref{fig:silicate_absorption_comparison_a} by red and blue curves, with correspondingly shaded one-sigma uncertainty regions. Figure~\ref{fig:silicate_absorption_comparison_a} also includes the weighted average for continuum-divided silicate-poor mid-L spectra (in black) as an estimate of the accuracy of the continuum removal process. Because the continuum in the silicate absorption region was determined from the silicate-poor spectra, the continuum-divided weighted average of the silicate-poor spectra scatters around unity. 

\section{Results: Gravity-dependent Silicate Absorption} 
\label{sec:results}

Figure~\ref{fig:silicate_absorption_comparison_a} shows that young silicate-rich mid-L spectra have a distinct absorption profile compared to old silicate-rich mid-L spectra. We observe three main differences.

\textbf{1.\ Silicate absorption strength peaks at a 0.4~$\mu$m longer wavelength in young mid-L dwarfs.} While the silicate absorption profiles for both the young and the old silicate-rich mid-L subsets were normalised to the same flux at 9.3~$\mu$m, Figure~\ref{fig:silicate_absorption_comparison_a} shows that the central wavelengths at which each of the profiles is the deepest are different. We estimate these central wavelengths by fitting a third-order polynomial to each continuum-divided weighted-average silicate-rich spectrum over the 8.0--11.5~$\mu$m wavelength range. We find that silicate absorption peaks at 9.5$\pm$0.1~$\mu$m in young silicate-rich mid-L spectra and at 9.1$\pm$0.1~$\mu$m in old silicate-rich mid-L spectra, as shown by the red vertical dashed line and the blue vertical dotted line in Figure~\ref{fig:silicate_absorption_comparison_a}. We estimate the central wavelength uncertainties from $10^4$ Monte Carlo realizations of each scaled and averaged spectrum. We consider the flux at each wavelength as a Gaussian random variate centred on the weighted average flux, and a standard deviation equal to the flux uncertainty. The central wavelength was measured in each realisation as explained above. The quoted uncertainty on the central wavelength is the 68\% confidence interval on the central wavelengths from the Monte Carlo simulation.

\textbf{2.\ Silicate absorption in young mid-L dwarfs is nearly twice as broad.} When scaled to the same absorption depth at 9.3~$\mu$m, the equivalent width of the silicate absorption in young mid-L dwarfs is 1.0$\pm$0.1~$\mu$m, whereas for the old mid-L dwarfs it is 0.6$\pm$0.1~$\mu$m. We estimate the uncertainties on the equivalent widths from the same Monte Carlo simulation used for the uncertainties on the central wavelengths of the silicate absorption above. The absorption feature in young mid-L dwarfs spans almost the entire plotted range in Figure~\ref{fig:silicate_absorption_comparison_a}, from $\sim$7.5~$\mu$m to $>$13.0~$\mu$m. In old mid-L dwarfs the silicate absorption strength tapers off beyond $\approx$11.5~$\mu$m. Low-level silicate absorption in old mid-L dwarfs may still be present at $>$11.5~$\mu$m, as in young mid-L dwarfs. However, even in our weighted-averaged spectra the difference from the silicate-poor 11.5--13.0~$\mu$m continuum is marginal.

\textbf{3.\ Silicate absorption in young mid-L dwarfs is more asymmetric towards longer wavelengths.}
Young mid-L dwarfs with silicate-rich spectra have a flatter silicate absorption profile for wavelengths longer than the feature peak compared to a sharper absorption profile in old mid-L dwarfs with silicate-rich spectra. The difference is most pronounced in the 10--12~$\mu$m wavelength range, where the scaled and averaged silicate absorption profile in young mid-L dwarfs is significantly stronger than in old mid-L dwarfs. Conversely, over 7.5--9.0~$\mu$m the absorption profiles are fully consistent with each other. 

Overall, the silicate absorption profile in young ($\sim$20--150~Myr; $\log g{\lesssim}4.5$) silicate-rich mid-L spectra is redder, broader, and more asymmetric than in old ($\gtrsim$500~Myr; $\log g{\gtrsim} 5.0$) silicate-rich mid-L spectra. This constitutes the first observational evidence for a surface gravity dependence of the mid-infrared silicate absorption in ultracool dwarf spectra.

\section{Discussion}
\label{sec:discussion}

Prior to \citet{Suarez-Metchev2022}, the existing Spitzer IRS spectra of the six L dwarfs inferred to show silicate absorption \citep{Roellig_etal2004,Cushing_etal2006,Looper_etal2008b} offered the only opportunity to analyse the composition of silicates in brown dwarf atmospheres. The consensus picture from that sample is that the mid-infrared silicate absorption in L dwarfs is dominated by $\sim$0.1--1.0~$\mu$m amorphous enstatite or quartz grains that reside in optically thin cloud slabs near the top of the atmosphere, at temperatures of $\sim$1100~K and pressures of $\lesssim$0.01~bar (\citealt{Burningham_etal2021, Luna-Morley2021}, and references therein). This silicate cloud layer may overlay a deeper iron cloud deck \citep{Burningham_etal2021}. A similar multi-layer scenario, comprising a patchy forsterite cloud above a deeper, optically thick iron deck, represents well the silicate absorption inferred in two young early-T dwarfs \citep{Vos_etal2023}.

The mid-infrared atmospheric modelling of silicates presented in \citet{Luna-Morley2021} is the broadest to date. They model five of the six L3.5--L6 dwarfs with previously detected silicate absorption from Spitzer IRS. By virtue of the similarity in spectral types, the best-fit models in \citet{Luna-Morley2021} all have similar fundamental parameters: four objects are inferred to have an effective temperature of $T_{\rm eff}=1800$~K and a surface gravity of $\log g=5$ and one has $T_{\rm eff}=1700$~K and $\log g=4$. The one object (2MASS J1507476$-$162738; L5) with a low-gravity best-fit model for the mid-infrared spectrum is unlikely to be young, however. It has no spectroscopic signatures of youth in $R{\sim}900$ optical \citep{Reid_etal2000} or $R{\sim}2000$ near-infrared \citep{Rayner_etal2009} spectra, and the BANYAN $\Sigma$ algorithm yields a 99.9\% probability of being an old field object. \citet{Filippazzo_etal2015} find a semi-empirical surface gravity of $\log g{=}5.18$, based on an assumed $>$500~Myr age and evolutionary models. As we have shown in \citet{Suarez_etal2021a}, $R{\sim}100$ Spitzer IRS spectra by themselves, as analysed in \citet{Luna-Morley2021}, may yield substantially less accurate results for effective temperature and surface gravity when fit by photospheric models compared to $R{\approx}2300$ near-infrared spectra. Hence this object has a likely higher gravity than inferred from the model best fit to the IRS spectrum in \citet{Luna-Morley2021}. Overall, the small number of previously known silicate-rich L-dwarf IRS spectra analysed in \citet{Luna-Morley2021} and the similarity in effective temperatures and surface gravities, do not allow a differentiation among the mid-infrared silicate absorption feature as a function of fundamental parameters. 

However, \citet{Luna-Morley2021} do produce detailed predictions for the spectroscopic signatures of  silicate condensates that can be compared to anticipated spectra from JWST. The recognition of a much larger sample of L dwarfs with detectable silicate absorption in Spitzer IRS spectra \citep{Suarez-Metchev2022}, and the gravity-dependent sub-sampling that we have performed here, allow a first look at systematic effects as a function of surface gravity.

\subsection{Dust Clouds in Young Atmospheres are Dominated by Heavier Grains}
\label{sec:heavier_grains}

We use the \citet{Luna-Morley2021} predictions to extract differences in silicate composition, grain size, or crystallinity that match the observed spectroscopic differences between low- and high-surface gravity mid-L dwarfs (Section~\ref{sec:results}). Specifically, we use the predictions for the mid-infrared silicate flux ratio $F_{\rm ratio}$ from Figure~7 in \citet{Luna-Morley2021}, wherein $F_{\rm ratio}$ is defined as the wavelength-dependent ratio of the grey-cloud ($F_{\rm grey}$) to the condensate-cloud ($F_{\rm cloud}$) model atmosphere. In so far as the grey-cloud model is representative of our featureless silicate-poor mid-L spectra, the \citet{Luna-Morley2021} $F_{\rm ratio}$ metric is similar to the inverse of our continuum-divided spectra (Figure~\ref{fig:silicate_absorption_comparison_a}). We plot this inverse in Figure~\ref{fig:silicate_absorption_comparison_b}. The only difference is that in \citet{Luna-Morley2021} the grey-cloud and condensate-cloud models are not scaled to each other in the silicate-free wavelength regions, while we have scaled both silicate-poor and silicate-rich mid-L spectra to the same average continuum (Figure~\ref{fig:silicate_absorption}). 

A comparison between our Figure~\ref{fig:silicate_absorption_comparison_b} and Figure 7 of \citet{Luna-Morley2021} reveals that the redder and more rounded 9.5~$\mu$m silicate absorption profile in young low-gravity mid-L dwarfs is likely caused by 1~$\mu$m amorphous magnesium- and iron-bearing pyroxene (Mg$_x$Fe$_{1-x}$SiO$_3$) grains. Pyroxene is the only iron-bearing silicate modelled in \citet{Luna-Morley2021} with a profile as asymmetric as the observed one, including a steeper slope over 8.0--9.5~$\mu$m and a flatter slope over 9.5--13~$\mu$m. An admixture of larger ($>$1~$\mu$m) amorphous enstatite grains is also possible, given the inverse dependence of peak wavelength on dominant grain size \citep{Luna-Morley2021}. However, the grains could not be as large as 10~$\mu$m, as they would  lack the characteristic absorption features and would instead behave as a source of grey opacity \citep{Cushing_etal2006,Luna-Morley2021}.

Conversely, the $\approx$9.1~$\mu$m peak and the narrower profile of silicate absorption in old high-gravity mid-L dwarfs points to the presence of $\lesssim$0.1~$\mu$m amorphous enstatite (MgSiO$_3$) or SiO grains. Figure 7 of \citet{Luna-Morley2021} shows 0.1~$\mu$m amorphous enstatite grains peaking in absorption strength around 9.2~$\mu$m in a 1800~K atmosphere, vs.\ a peak at 9.4~$\mu$m for larger 1~$\mu$m grains. One-micron amorphous SiO grains exhibit a similar absorption peak and profile as 1~$\mu$m amorphous enstatite grains, and are also expected to peak closer to 9.1~$\mu$m if smaller in size. Contributions from other species are likely secondary.  Thus, a contribution from amorphous or crystalline quartz (SiO$_2$), which peaks in strength at $\approx$8.8~$\mu$m is possible, but not likely, given the sharpness of the SiO$_2$ absorption feature \citep{Luna-Morley2021}. A more accurate determination of the silicate and iron content and grain size of dust clouds in brown dwarf atmospheres awaits JWST MIRI observations, which will bring broader wavelength coverage at higher SNR and spectral resolution. 

While the comparison to model predictions is not conclusive about the composition and size of the silicate grains in low- vs.\ high-gravity mid-L dwarfs, there is one consistent difference between the two. The types of grains that fit better the redder and broader silicate absorption observed in young low-gravity mid-L dwarfs are heavier than the grains that fit better the bluer and narrower silicate absorption profiles of old high-gravity mid-L dwarfs. 

\subsection{Lower-gravity Ultracool Atmospheres Have Lower Sedimentation Efficiency}
\label{sec:delayed_sedimentation}

The preceding spectroscopic analysis established that silicate grains in low-gravity atmospheres are both larger ($\sim$1~$\mu$m) and composed of a heavier molecule (Mg$_x$Fe$_{1-x}$SiO$_3$) compared to silicates in high-gravity atmospheres, which have sizes of $\lesssim$0.1~$\mu$m and are dominated by MgSiO$_3$ or SiO. For example, for a typical $x=0.4$ pyroxene (Mg$_{0.4}$Fe$_{0.6}$SiO$_3$) has a molecular weight of 23.9 g~mol$^{-1}$ per constituent atom, which is heavier than enstatite (MgSiO$_3$, 20.1 g~mol$^{-1}$~atom$^{-1}$) or SiO (22.0 g~mol$^{-1}$~atom$^{-1}$). Iron-rich pyroxene condensates are also more refractory than enstatite or SiO condensates, and so settle at lower altitudes in brown dwarf atmospheres (e.g., \citealt{Ackerman-Marley2001,Cooper_etal2003,Lodders2003,Woitke_etal2020}). We have thus found spectroscopic evidence that the settling of dust condensates in ultracool atmospheres depends on surface gravity.

This confirms the very well-established theoretical expectation that low-gravity atmospheres have lower sedimentation efficiencies \citep{Ackerman-Marley2001,Cooper_etal2003}, and that correspondingly the settling of dust clouds in low-gravity atmospheres happens at lower temperatures \citep{Stephens_etal2009,Madhusudhan_etal2011,Marley_etal2012,Charnay_etal2018}, leading to dustier atmospheres in young objects.

\clearpage
\begin{landscape}
\begin{table}
\begin{center}
\caption{Sub-samples of mid-L dwarfs with silicate-rich and silicate-poor spectra.}
  \small
  \label{tab:dwarfs}
  \begin{threeparttable}
	\begin{tabular}{lccccccc}
	\toprule
	Designation                         & Short Name  & SpT$_{\textrm{opt}}$ & Ref.                                & SpT$_{\textrm{IR}}$ & Ref.                         & Silicate Index & $\Delta$Si$^{\rm a}$ \\
    \midrule
	\multicolumn{8}{c}{\textbf{Young silicate-rich mid-L spectra}} \\
	2MASS J03552337+1133437$^{\rm b}$   & 0355+1133   & L5$\gamma$           & \citet{Cruz_etal2009}               & L4$\gamma$      & \citet{BardalezGagliuffi_etal2019}       & 1.29$\pm$0.02  & 0.19$\pm$0.02        \\
	2MASS J05012406$-$0010452$^{\rm c}$ & 0501$-$0010 & L4$\gamma$           & \citet{Cruz_etal2009}               & L4$\gamma$          & \citet{Gagne_etal2015}       & 1.35$\pm$0.08  & 0.20$\pm$0.08        \\
	2MASS J14252798$-$3650229$^{\rm b}$ & 1425$-$3650 & L3                   & \citet{Reid_etal2008}               & L4$\gamma$          & \citet{Gagne_etal2015}       & 1.32$\pm$0.04  & 0.20$\pm$0.04        \\
	2MASS J22443167+2043433$^{\rm b}$   & 2244+2043   & L6.5                 & \citet{Kirkpatrick_etal2008}        & L6--L8$\gamma$      & \citet{Gagne_etal2015}       & 1.37$\pm$0.07  & 0.38$\pm$0.07        \\
	\multicolumn{8}{c}{\textbf{Old silicate-rich mid-L spectra}} \\                                                                                                                                             
	2MASS J00043484$-$4044058           & 0004$-$4044 & L5                   & \citet{Kirkpatrick_etal2001}        & L4.5                & \citet{Knapp_etal2004}       & 1.36$\pm$0.11  & 0.25$\pm$0.11        \\
	2MASS J01443536$-$0716142           & 0144$-$0716 & L5                   & \citet{Liebert_etal2003}            & L5                  & \citet{Marocco_etal2013}     & 1.29$\pm$0.05  & 0.18$\pm$0.05        \\
	2MASS J09201223+3517429$^{\rm d}$   & 0920+3517   & L6.5                 & \citet{Kirkpatrick_etal2000}        & T0p                 & \citet{Burgasser_etal2006b}  & 1.31$\pm$0.14  & 0.32$\pm$0.14        \\
	2MASS J12392727+5515371             & 1239+5515   & L5                   & \citet{Kirkpatrick_etal2000}        & L5                  & \citet{Schneider_etal2014}   & 1.48$\pm$0.12  & 0.37$\pm$0.12        \\
	2MASS J14482563+1031590             & 1448+1031   & L4                   & \citet{Wilson_etal2003b}            & L5.5                & \citet{Schneider_etal2014}   & 1.26$\pm$0.05  & 0.10$\pm$0.05        \\
	2MASS J17534518$-$6559559           & 1753$-$6559 & L4                   & \citet{Reid_etal2008}               & L4                  & \citet{Marocco_etal2013}     & 1.37$\pm$0.09  & 0.22$\pm$0.09        \\
	2MASS J18212815+1414010             & 1821+1414   & L4.5                 & \citet{Looper_etal2008b}            & L5p                 & \citet{Kirkpatrick_etal2010} & 1.25$\pm$0.01  & 0.11$\pm$0.01        \\
	2MASS J21481633+4003594             & 2148+4003   & L6                   & \citet{Looper_etal2008b}            & L6.5p               & \citet{Kirkpatrick_etal2010} & 1.44$\pm$0.01  & 0.40$\pm$0.01        \\
	2MASS J22244381$-$0158521           & 2224$-$0158 & L4.5                 & \citet{Kirkpatrick_etal2000}        & L3.5                & \citet{Knapp_etal2004}       & 1.38$\pm$0.06  & 0.24$\pm$0.06        \\
	\multicolumn{8}{c}{\textbf{Silicate-poor mid-L spectra}} \\                                                                                                                          
	2MASS J07003664+3157266             & 0700+3157   & L3.5                 & \citet{Thorstensen-Kirkpatrick2003} & \nodata             & \nodata                      & 1.02$\pm$0.04  & $-0.11\pm0.04$       \\
	2MASS J09083803+5032088             & 0908+5032   & L5                   & \citet{Cruz_etal2003}               & L9                  & \citet{Knapp_etal2004}       & 0.89$\pm$0.04  & $-0.21\pm0.04$       \\
	2MASS J10365305$-$3441380           & 1036$-$3441 & L6                   & \citet{Gizis2002}                   & L6.5                & \citet{Burgasser_etal2010b}  & 0.83$\pm$0.05  & $-0.20\pm0.05$       \\
	2MASS J11263991$-$5003550           & 1126$-$5003 & L4.5                 & \citet{Burgasser_etal2008b}         & L6.5p               & \citet{Burgasser_etal2008b}  & 1.01$\pm$0.04  & $-0.13\pm0.04$       \\
	2MASS J13314894$-$0116500           & 1331$-$0116 & L6                   & \citet{Hawley_etal2002}             & L6                  & \citet{Marocco_etal2015}     & 0.86$\pm$0.06  & $-0.18\pm0.06$       \\
	2MASS J15150083+4847416             & 1515+4847   & L6                   & \citet{Cruz_etal2007}               & L6.5                & \citet{Cruz_etal2003}        & 0.88$\pm$0.04  & $-0.15\pm0.04$       \\
	2MASS J16262034+3925190             & 1626+3925   & sdL4                 & \citet{Burgasser_etal2007}          & sdL                 & \citet{Burgasser2004c}       & 0.95$\pm$0.06  & $-0.20\pm0.06$       \\
	\bottomrule
	\end{tabular}
	\begin{tablenotes}[para,flushleft]
	  $^{\rm a}$Silicate index excess defined as the difference between the measured silicate index value for each dwarf and the median index for its spectral type (Section~\ref{sec:sample_selection} and Figure~\ref{fig:silicate_index_vs_spt_a}).\\
	  $^{\rm b}$Member of the 110--150~Myr-old \citep{Luhman_etal2005b,Barenfeld_etal2013} AB Doradus moving group. References are \citet{Liu_etal2013,Gagne_etal2015,Vos_etal2018} for 0355+1133, 1425$-$3650, and 2244+2043, respectively.\\
	  $^{\rm c}$Member of either the Columba or Carina moving groups \citep{Gagne_etal2015}, which are coeval groups at 20--40~Myr \citep{Torres_etal2008}.\\
	  $^{\rm d}$Close binary system with the individual components having near-infrared spectral types of L5.5 and L9 \citep{Dupuy-Liu2012}. The binary is not resolved in the Spitzer observations and, therefore, the IRS spectrum is likely dominated by the brighter L5.5 dwarf.
	\end{tablenotes}
  \end{threeparttable}
\end{center}
\end{table}
\end{landscape}

\clearpage
\twocolumn

\section{Conclusion}
\label{sec:summary}

We have established that the mid-infrared silicate absorption in the spectra of mid-L dwarfs is sensitive to surface gravity (Section~\ref{sec:results}). It is broader, redder, and more asymmetric for young $\log g{\lesssim} 4.5$ mid-L dwarfs than for their older $\log g{\gtrsim} 5.0$ counterparts. Comparisons to theoretical spectra reveal that the silicate grains in young mid-L dwarfs are heavier (Section~\ref{sec:heavier_grains}). They are larger ($\sim$1~$\mu$m) and dominated by silicates that are more refractory (magnesium- and iron-bearing pyroxene) than grains in the atmospheres of older mid-L dwarfs ($\lesssim$0.1~$\mu$m in size and dominated by enstatite or SiO). This result directly confirms that dust cloud sedimentation in low-gravity substellar atmospheres is suppressed (Section~\ref{sec:delayed_sedimentation}), as already anticipated by their redder optical and near-infrared photospheric colours \citep{Kirkpatrick2005,Filippazzo_etal2015} and by their enhanced photometric variability (e.g., \citealt{Metchev_etal2015,Biller_etal2015}). The surface gravity dependence of the condensate physics and chemistry enhances the empirical context for JWST studies of brown dwarf and exoplanetary atmosphere.

\section*{Acknowledgements}
We thank the referee Denise C. Stephens for a helpful report that significantly improved the quality of this paper. We thank Marina Gemma for discussions and suggestions on the use of appropriate nomenclature for dust grains. G.S. acknowledges support from NSF award \#2009177. Support for this work was provided by the Canadian Space Agency under the Flights and Fieldwork for the Advancement of Science and Technology programme (reference no.\ 19FAWESB40). The spectra presented here are obtained entirely with the Spitzer Space Telescope, which was operated by the Jet Propulsion Laboratory, California Institute of Technology under a contract with NASA.

\section*{Data Availability}

The spectra used in this analysis form part of the uniformly reprocessed set of Spitzer IRS spectra of 113 M5--T9 dwarfs published and available in \citet{Suarez-Metchev2022} and on Genaro Su\'arez' homepage\footnote{\url{https://gsuarezcastro.wixsite.com/gsuarez/research}}.



\bibliographystyle{mnras}
\bibliography{mybib_Suarez} 







\bsp	
\label{lastpage}
\end{document}